# Phase Contrast technique: a broader approach.

David Sanchez-de-la-Llave, M. David Iturbe-Castillo, and Carlos G Trevino-Palacios

Instituto Nacional de Astrofísica, Optica y Electronica

Apdo Postal 51 y 216, Puebla Pue 72000 Mexico

The treatment found on most general optics textbooks related to the phase contrast technique imposes limitations on the filter phase and object phase variations in order to mathematically explain it in a simple manner. We consider that this simplified treatment can be misleading in terms of the concept the student may develop and also about the potential applications of the phase contrast technique. In this paper we describe a broader and yet simple explanation of the phase contrast process, creating a parallelism between optical image processing and interferometry.

## INTRODUCTION

Phase objects alter the phase but not the amplitude of the incident light, for example thermal variations in air. Typical detection systems are ineffective in the analysis of such objects since the eye, CCD cameras, photomultipliers and other light detection devices are sensitive only to variations in intensity and not phase.

Phase objects are of variations in thickness, refractive index or combinations of these, leading to a variation in optical thickness of the object. The method suggested by Zernike[1], known as Phase Contrast Technique, for which he was awarded the Nobel

Prize in 1953, converts phase modulation caused by the object into amplitude modulation. In the case of weak phase objects, this conversion is performed such that phase variations map linearly to amplitude variation in the image field thus rendering an image depicting the variation of optical thickness of a phase object.

In most classical textbooks related to general optics [2,3,4,5,6,7,8,9] the treatment found on the phase contrast technique is based on two important conditions. The first is related to the filter phase, always set to $\pi/2$ rad. The second is concerned with the object phase variations allowed, always limited to be smaller than 0.1 rad. These conditions are required in order to describe the system in a simplified manner. However, we consider that the simplified treatment can be reformulated in terms of the concept that the student may develop about the technique and the potential applications of the phase contrast technique. In this paper we describe a broader and yet simple explanation of the phase contrast process. The description presented herein is adequate for courses in which optical information processing is the main subject. This alternative way of teaching phase contrast gives the student a better understanding of the topic, provides a parallelism with interferometry and opens a broader view on its applications.

We begin by giving a review on the phase contrast method as presented in general optics textbooks. Then we present an alternative approach, in which no limitations are imposed on either the filter phase or the amplitude of the object phase variations. Finally some conclusions are drawn.

## Simplified treatment

The phase contrast technique is employed to retrieve the phase information. The most widely used textbooks on general optics devote some part of a chapter to this problem. Two important considerations are made. One is that the filter phase, which only alters the object's zero frequency spectral component, is restricted to a value of $\pm\pi/2$ rad. The second one is that the object phase variations, *Φ(x,y)*, are small (less than 0.1 rad). The phase function, *exp[i(Φ(x,y))]*, where the average phase shift through the object has been omitted, is expressed as a Taylor series. Using the first assumption, the object phase function expansion is left with its first two expansion terms (*1+ iΦ(x,y)*). Since *Φ(x,y)* represents variations around the average phase shift, it has no energy at the zero frequency spectral component. Therefore, the filtering process only alters the constant term of the object phase function approximation. Consequently, at the image plane both the constant term and the phase distribution *Φ(x,y)* are imaginary functions (*i + iΦ(x,y)*). Calculating the intensity distribution at the image plane, where the term *Φ²(x,y)* has been neglected based on the assumption that *Φ(x,y)* is small, yields *I(x,y)=1±2 Φ(x,y).* Using this simplified explanation a linear relationship between intensity at the image plane and the object phase distribution is found.

## General treatment

The phase contrast method can be explained with a treatment without the two abovementioned restrictive conditions. For simplicity, we present a one-dimensional treatment, however the generalization to two-dimensional structures is straightforward.

We start with a general phase filter which affects only the zero frequency and leaves unaltered the rest of the spectrum, given by

$$H(\upsilon) = 1 + \lim_{w \to 0} rect\left(\frac{\upsilon}{w}\right)(exp[i\alpha] - 1), \qquad (1)$$

where *α* is a constant, and *rect* is the rectangle function give by:

$$rect\left(\frac{\upsilon}{w}\right) = \begin{cases} 1 & for\ |w/2| \geq \upsilon \\ 0 & otherwise \end{cases}. \qquad (2)$$

This simple form of the filter function states that the phase filter can be interpreted as the sum of two filters: one that leaves the object spectrum unaltered and one that only alters the zero frequency component and blocks the rest of the spectrum.

To implement the phase contrast technique consider a 4f optical system with a monochromatic plane wave with unitary amplitude used to illuminate the phase object located at the input plane. This is an imaging system with unitary magnification and a Fourier plane where the input object Fourier transform and the field distribution are proportionally related (figure 1).

Considering a general phase object described by *f(x)=exp[iΘ(x)]*, where *Θ(x)* can take any value, the field distribution at the intermediate Fourier plane produced by the first lens and the phase filter is given by

$$F(\upsilon)H(\upsilon) \propto F(\upsilon) + \lim_{w \to 0} rect\left(\frac{\upsilon}{w}\right)(exp[i\alpha] - 1)F(0), \qquad (2)$$

where *F(υ)* is the Fourier transform of *f(x)*.

After a second Fourier transformation, produced by the second lens, the field distribution at the image plane is

$$o(x') \propto f(x') + F_0[\exp[i\alpha] - 1], \quad (3)$$

where $F_0 = |F_0| \exp(i\theta_{F_0}) = F(0)$.

Expression (3) is the superposition of two terms: the information of the phase object (first term) and a plane wave from the zero frequency (second term). We interpret these as two coherent sources. The second lens and the detection process makes them interfere, therefore one could think of the phase contrast system as an interferometer in which the phase filter acts as a point source. This is known as a common path interferometer[10]. The intensity distribution at the image plane is

$$|o(x')|^2 \propto \left[1 + 2|F_0|^2(1-\cos\alpha)\right] + 2|F_0|[2(1-\cos\alpha)]^{1/2} \cos\left(\tan^{-1}\left(\frac{\sin\alpha}{\cos\alpha - 1}\right) + \theta_{F_0} - \Theta(x')\right), \quad (4)$$

which is proportional to a constant term $1 + 2|F_0|^2(1-\cos\alpha)$ added to a cosine function with amplitude $2|F_0|[2(1-\cos\alpha)]^{1/2}$ and argument $\tan^{-1}[\sin\alpha/(\cos\alpha - 1)] + \theta_{F_0} - \Theta(x')$. The cosine function can be rewritten as

$$\cos\left(\tan^{-1}\left(\frac{\sin\alpha}{\cos\alpha - 1}\right) + \theta_{F_0} - \Theta(x')\right) = \begin{cases} -\sin\left(\frac{|\alpha|}{2} - \theta_{F_0} + \Theta(x')\right), & -\pi \leq \alpha < 0 \\ -\sin\left(\frac{|\alpha|}{2} + \theta_{F_0} - \Theta(x')\right), & 0 < \alpha < \pi \end{cases}. \quad (5)$$

In order to analyze the expression for the field distribution at the image plane (4) without restricting the treatment to the textbook small phase variations on the object or a filter phase equal to $\pm\pi/2$ rad, we consider the cases of small and large object phase variations without restriction on the phase filter value.

**Small phase variations**

Let us first consider an object with small phase variations $\Phi(x,y)$ around an average constant phase. The average constant phase represents a uniform phase delay throughout the object domain. If we are not interested in measuring absolute values for the phase variations, then the constant phase delay can be neglected. Also, the term $F_0$ can be made equal to 1 and, as such, it does not affect the argument of the cosine function, i.e., $\theta_{Fo}=0$ in expressions (4) and (5). The filter phase $\alpha$ plays an important role regarding the linear relationship between intensity and the object phase distribution, as it can be observed in expressions (4) and (5). A graph of the sine function with argument α/2 between 0 and π/2 rad (figure 2) aids to show that a filter phase different from α =±π/2 rad can be employed and still get a linear relationship between the interferogram intensity and the small phase distribution. In order to show that expression (4) reduces to the linear relationship described in textbooks we consider the example of a filter phase equal to π/2 rad. To demonstrate that a linear relationship can be obtained for other filter phases we consider the case of a filter phase equal to π/4 rad.

*π/2 rad phase filter (Zernike filter)*

For the general case of small phase variations and a positive filter phase, expression (4) can be written as:

$$|o(x')|^2 \propto [1+2(1-\cos\alpha)] - 2[2(1-\cos\alpha)]^{1/2} \sin\left(\frac{|\alpha|}{2} - \Phi(x')\right). \qquad (6)$$

Considering a the phase filter ($\alpha$) equal to π/2 rad, equation (6) is

$$|o(x')|^2 \propto 3 - 2[2]^{1/2} \sin\left(\frac{\pi}{4} - \Phi(x')\right) = 3 - 2[2]^{1/2}\left([2]^{-1/2} \cos(\Phi(x')) - [2]^{-1/2} \sin(\Phi(x'))\right). \quad (7)$$

Since $\Phi(x)$ is small, $cos(\Phi(x))\approx 1$ and $sin(\Phi(x)) \approx \Phi(x)$, therefore equation (7) can be rewritten as

$$|o(x')|^2 \propto 1 + 2\Phi(x'). \quad (8)$$

This is the same result using the simplified analysis of the system. The analysis is in agreement with the textbook description of the phase contrast technique aforementioned, but provides a richer interpretation.

### π/4 rad phase filter

If we consider a phase filter different from π/2 the analysis is similar. In the case of α=π/4 rad, equation (6) becomes

$$|o(x')|^2 \propto 1.5858 - 1.5307\left(\sin\left(\frac{\pi}{8}\right)\cos(\Phi(x')) - \cos\left(\frac{\pi}{8}\right)\sin(\Phi(x'))\right) \approx 1 + 1.4142\ \Phi(x'), \quad (9)$$

where a linear relationship between interferogram intensities and phase variations is obtained under the small object phase approximation.

### Arbitrary phase filter

Any filter phase different from π/2 rad can produce a linear relationship between the phase distribution and the interferogram intensity. However, there is a tradeoff involved in modifying the π/2 rad Zernike filter phase.

Taking $|\Phi_{max}|$ as the object's maximum phase variation, considered small, we calculate the extreme intensities using (4) and (5); these are $I_{max}$ and $I_{min}$, the maximum and minimum interferogram intensities, respectively.

$$I_{max} \propto 1 + 2(1 - \cos\alpha) - 2[2(1 - \cos\alpha)]^{1/2}\left[\sin\left(\frac{\alpha}{2}\right) + \cos\left(\frac{\alpha}{2}\right)|\Phi_{max}|\right]$$
$$I_{min} \propto 1 + 2(1 - \cos\alpha) - 2[2(1 - \cos\alpha)]^{1/2}\left[\sin\left(\frac{\alpha}{2}\right) - \cos\left(\frac{\alpha}{2}\right)|\Phi_{max}|\right]$$
(10)

To evaluate the performance of the phase contrast apparatus we assess the image contrast defined as $C=(I_{max}-I_{min})/(I_{max}+I_{min})$. Using trigonometric identities and keeping in mind that $|\Phi_{max}|$ is small, the contrast is given by

$$C = 2|\Phi_{max}\sin(\alpha)|. \qquad (11)$$

The contrast reaches its maximum value for the π/2 rad filter phase (see figure 3). If we employ a phase filter larger than π/2 rad, the maximum phase variations under which a linear relationship could be obtained would be reduced and tend to zero as the phase filter approaches π rad (see figure 2).

**Large phase variations**

From figure 2, one might be tempted to think that a linear relationship could be obtained between larger phase variations and intensity by using a phase filter smaller than π/2 rad. However, for these larger phase variations the term $F_0$ can not be made equal to one and is rather object dependent. Therefore, although a linear relationship between phase distribution and intensity indeed exists, it can not be determined.

Hence, in general, when we do not restrict the object phase variations to be small a nonlinear relationship is obtained between phase variations and output intensity, just like in a classical interferometer. The term $\boldsymbol{F_0}$ turns into a complex number, inducing an unknown constant term in the argument of the cosine function. Nevertheless, numerical techniques can be applied to the resulting interferogram in order to retrieve the object phase distribution. Furthermore, the filter phase value is no longer restricted; it can be any number different from zero or $2\pi$ multiples. In this case, the filter phase plays an important role on the interferogram contrast.

**Conclusions**

The treatment found on most general optics textbooks imposes limitations on the filter phase and object phase variations in order to mathematically explain the phase contrast technique. Such limitations are justified when time is a constraint. However the more general approach we propose here can enhance the understanding the student may develop of such a system. The general treatment we propose requires some basic concepts regarding Fourier optics, which are also required in the classical treatment. Finally, we consider that learning the phase contrast technique with the general treatment presented here opens up a wider variety of applications in the students mind either for small of large phase variations objects and help in making an analogy between optical image processing and interferometry.

REFERENCES

1. F. Zernike, "Diffraction theory of knife-edge test and its improved form, the phase contrast," Mon. Not. Roy. Astr. Soc. **94**, 382-383 (1934).

2. Francis A. Jenkins and Harvey E. White, *Fundamental of Optics*, **4**th edition (MacGraw Hill, 1991), Chap. 28, p.602-604.

3. Miles V. Klein and Thomas E. Furtak, *Optics*, **2**nd edition (Wiley, 1986), Chap. 7, p.485-487.

4. Robert Guenther, *Modern Optics* (Wiley, 1990), Chap. 10, p.413-415.

5. Stephen G. Lipson, H. Lipson, and D. S. Tannhauser, *Optical Physics*, **3**rd edition (Cambridge University Press, 1995), Chap. 12, p.347-349.

6. Joseph W. Goodman, *Introduction to Fourier Optics* (McGraw Hill, 1996), Chap. 8, p.220-222.

7. Eugene Hecht, *Optics*, **3**rd edition (Addison Wesley, 1998), Chap. 13, p.611-615.

8. Max Born and Emil Wolf, *Principles of Optics*, **7**th edition (Cambridge University Press, 1999), Chap. 8, p.472-475.

9. Francis Graham-Smith and Terry A. King, *Optics and Photonics: an introduction* (Wiley, 2000), Vol. , Chap. 14, p.276-277.

10. Chirs S. Anderson, "Fringe visibility, irradiance and accuracy in common path interferometers for visualization of phase disturbances," Appl. Opt. **34**, 7474-7485 (1995).

**FIGURE CAPTION**

Figure 1. Unitary magnification 4f optical system. Two identical lenses L1 and L2 are placed at twice their focal length, f.

Figure 2. Weight function for the intensity distribution in equation (4). The classical Zernike case ($|\alpha| = \pi/2$) is mark by the arrow.

Figure 3. Contrast for filter phases between 0 and $\pi$. The maximum object phase variation ($|\Phi max|$) is set to 0.1 rad.

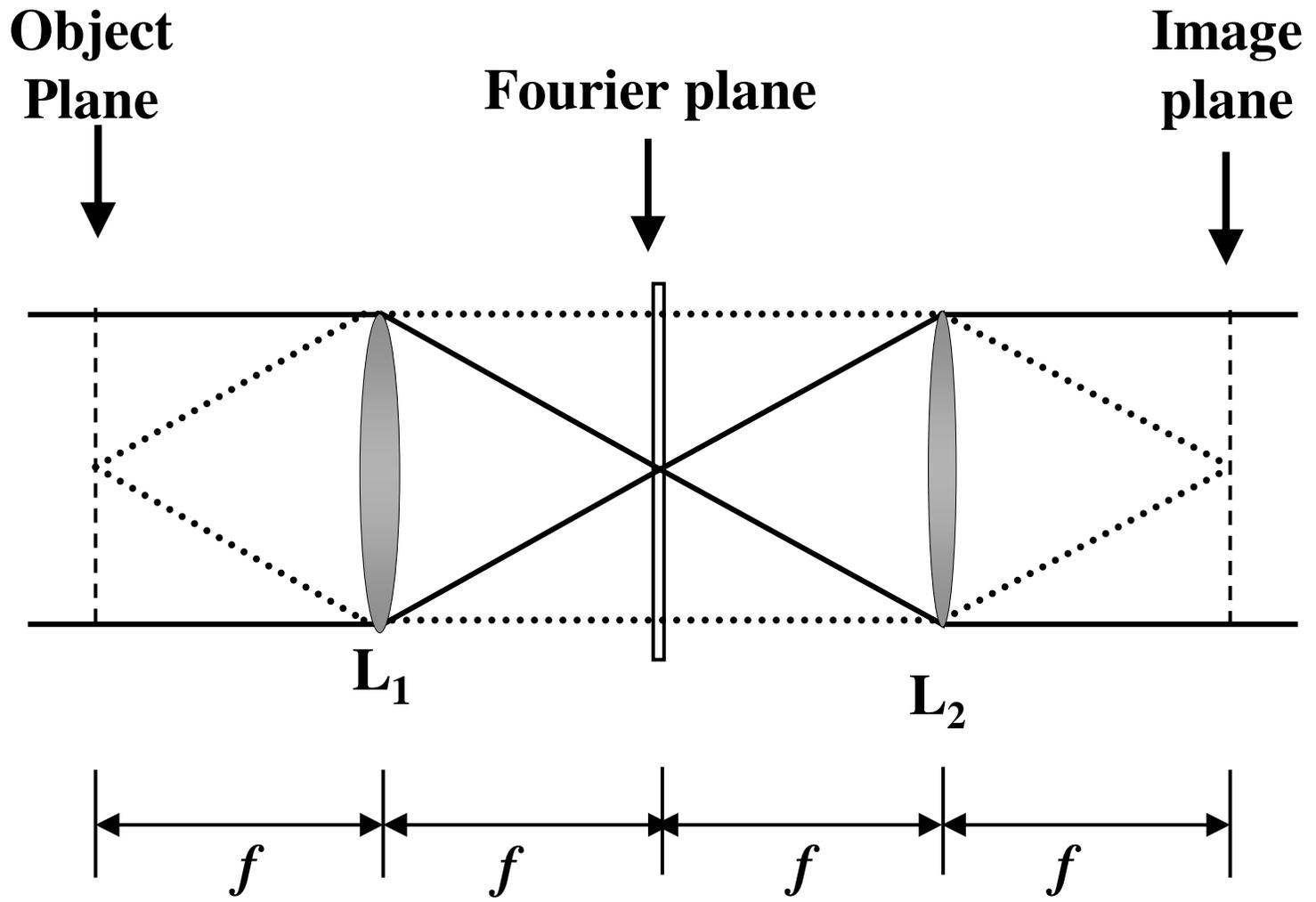

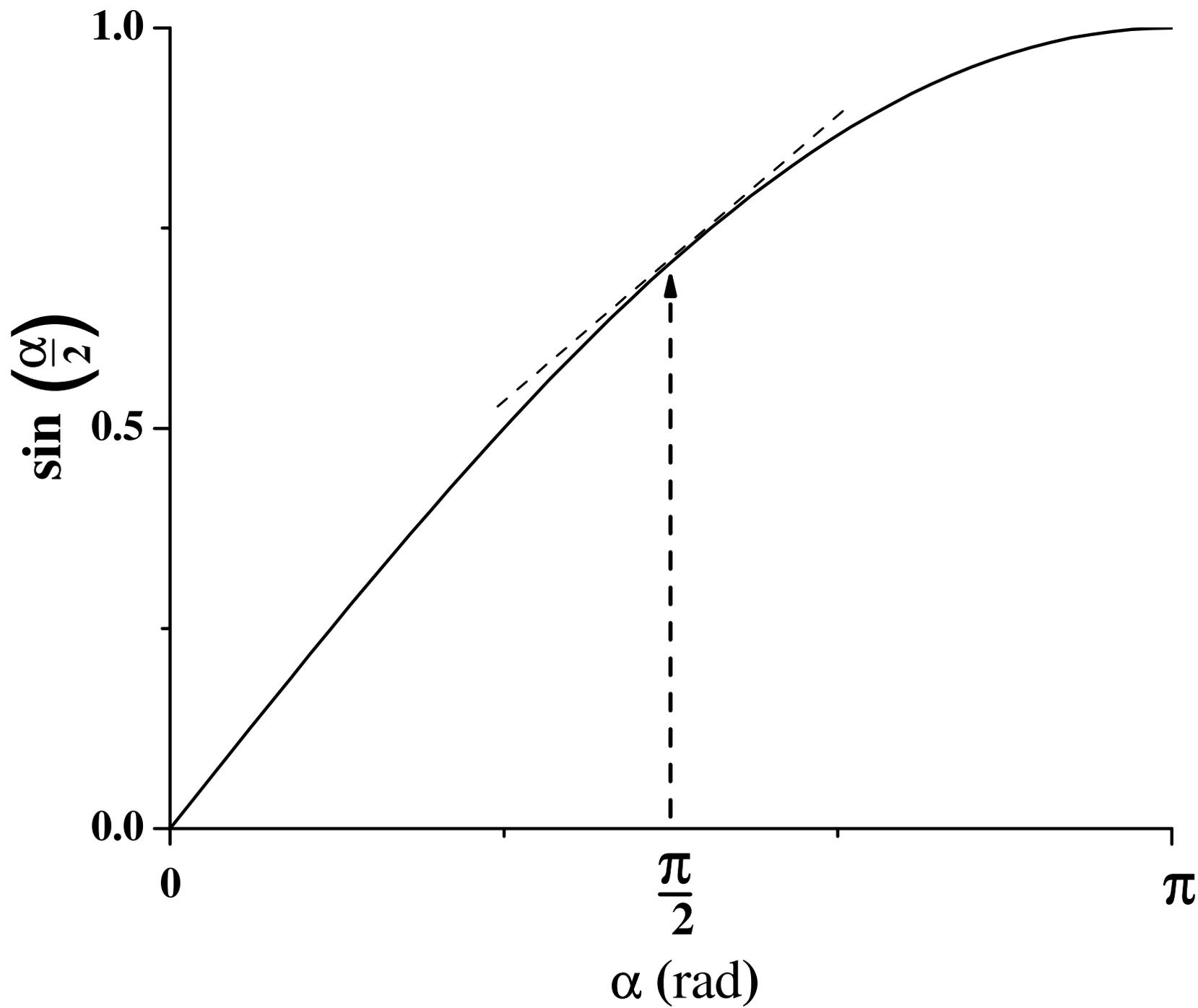

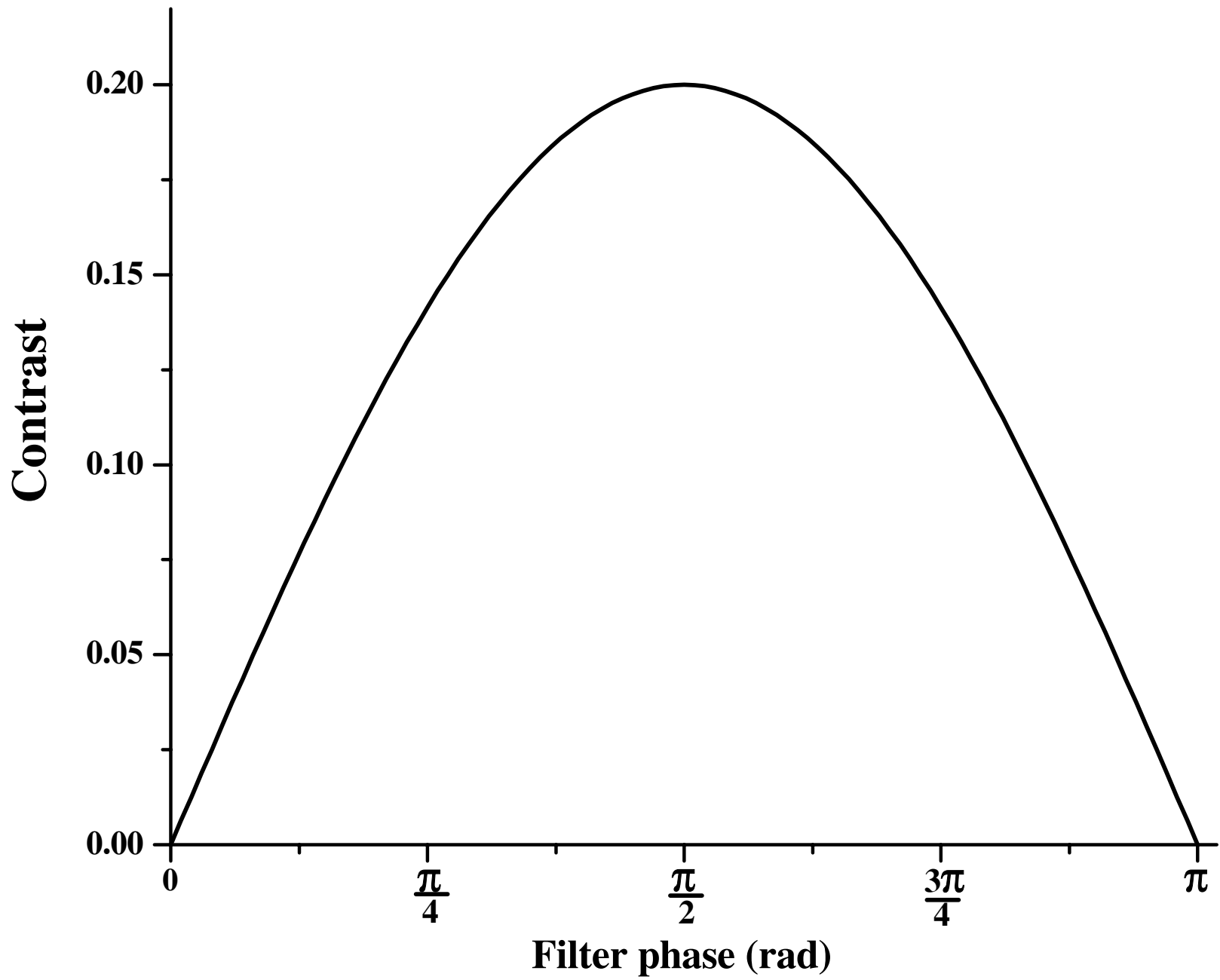